\documentclass[sort&compress,final,5p,twocolumn,fixfloat]{elsarticle}

\journal{Physics Letters B}

\usepackage{color}
\usepackage[dvipsnames]{xcolor}
\usepackage{bm}
\usepackage{amsmath}
\usepackage{amssymb}
\usepackage{booktabs}
\usepackage{float}
\usepackage[caption=false]{subfig} 
\usepackage{tabularx}
\usepackage{scalerel}
\usepackage{tikz}
\usetikzlibrary{svg.path}

\usepackage[hidelinks]{hyperref}
\usepackage[sort&compress, capitalise]{cleveref}
 \hypersetup{
   pdftitle={},%
   pdfauthor={},%
   pdfsubject={},%
   pdfkeywords={},%
   pdfstartview={},%
   bookmarksopen=true, breaklinks=true, debug=true, %
   colorlinks=true, linkcolor=red, citecolor=blue, urlcolor=blue
 }

\newcommand{\ie}{{\it i.e.~}}
\newcommand{\eg}{{\it e.g.~}}

\newcommand{\GeV}{\,\text{GeV}}
\newcommand{\itk}{\mathit{k}}
\newcommand{\wk}{\mathit{w}_\itk}
\newcommand{\bfa}{\boldsymbol{a}}
\newcommand{\bfy}{\boldsymbol{y}}

\newcommand{\bfk}{\boldsymbol{k}}

\usepackage[normalem]{ulem}

\newcommand*\oline[1]{%
   \vbox{%
     \hrule height 0.5pt
     \kern0.4ex
     \hbox{%
       \kern-0.15em
       \ifmmode#1\else\ensuremath{#1}\fi
       \kern-0.15em
     }
   }
}

\begin{document}

\title{Reweighting the Sivers function with jet data from STAR}

\date{\today}

\author[add1,add2]{M.~Boglione}
\ead{elena.boglione@to.infn.it}

\author[add3,add4]{U.~D'Alesio}
\ead{umberto.dalesio@ca.infn.it}

\author[add5]{C.~Flore\corref{cor1}}
\ead{carlo.flore@ijclab.in2p3.fr}
\cortext[cor1]{Corresponding author}

\author[add1,add2]{J.~O.~Gonzalez-Hernandez}
\ead{joseosvaldo.gonzalez@to.infn.it}

\author[add4]{F.~Murgia}
\ead{francesco.murgia@ca.infn.it}

\author[add6,add7]{A.~Prokudin}
\ead{prokudin@jlab.org}

\address[add1]{Dipartimento di Fisica Teorica, Universit\`a di Torino, Via P.~Giuria 1, Torino, I-10125, Italy}
\address[add2]{INFN, Sezione di Torino, Via P.~Giuria 1, Torino, I-10125, Italy}
\address[add3]{Dipartimento di Fisica, Universit\`a di Cagliari, Cittadella Universitaria, I-09042 Monserrato (CA), Italy}
\address[add4]{INFN, Sezione di Cagliari, Cittadella Universitaria, I-09042 Monserrato (CA), Italy}
\address[add5]{Universit\'e Paris-Saclay, CNRS, IJCLab, 91405 Orsay, France}
\address[add6]{Division of Science, Penn State University Berks, Reading, Pennsylvania 19610, USA}
\address[add7]{Theory Center, Jefferson Lab, 12000 Jefferson Avenue, Newport News, Virginia 23606, USA}

\begin{abstract}
The reweighting procedure that using Bayesian statistics incorporates the information contained in a new data set, without the need of re-fitting, is applied to the quark Sivers function extracted from Semi-Inclusive Deep Inelastic Scattering (SIDIS) data. We exploit the recently published single spin asymmetry data for the inclusive jet production in polarized $pp$ collisions from the STAR Collaboration at RHIC, which cover a much wider $x$ region compared to SIDIS measurements. The reweighting method is extended to the case of asymmetric errors and the results show a remarkable improvement of the knowledge of the quark Sivers function.
\end{abstract}


\maketitle

\section{\label{sec:intro}Introduction}

Single spin asymmetries (SSAs) in inclusive and semi-inclusive processes are an invaluable tool to deepen our knowledge of the internal structure of nucleons as well  the hadronization mechanism. Despite the wealth and richness of available data and an extensive theoretical effort carried out in the last decades, the understanding of their origin still represents a formidable challenge from the phenomenological point of view. Related to this issue, there is nowadays a general consensus that the three-dimensional (3D) picture of hadrons and the corresponding multi-parton correlations would lead to a better comprehension of the hadronic structure. 

The 3D hadron structure in momentum space is encoded in a new class of parton distribution and fragmentation functions, the so-called Transverse Momentum Dependent distributions and fragmentation functions (TMDs), which depend on the  collinear momentum fraction carried by the parton and its intrinsic  transverse  momentum. Among the eight leading-twist nucleon TMD distributions, the Sivers function~\cite{Sivers:1989cc,Sivers:1990fh} is one of the most studied and plays a seminal role. In fact, the Sivers distribution has several distinctive features: it is naively time-reversal odd, its existence is related to a nonvanishing parton orbital angular momentum and, more importantly, it is expected to be process dependent, having opposite signs in SIDIS and Drell-Yan (DY) processes~\cite{Collins:2002kn, Brodsky:2002rv}.

For two-scale processes, such as SIDIS and DY, where  $Q_2 \gg Q_1\sim \Lambda_{\rm QCD}$, TMD factorization~\cite{Ji:2004wu,Collins:2011zzd,Echevarria:2012js} is proven and SSAs are described in terms of convolutions of TMDs. These two scales are the virtuality of the exchanged boson and the transverse momentum of the observed hadron in SIDIS, and the invariant mass and transverse momentum of the lepton pair in DY. Therefore, such measurements are sensitive to the non perturbative transverse motion of bound partons in the nucleon encoded in TMDs. For processes with one characteristic perturbative scale, or $Q_1 \simeq Q_2 \gg  \Lambda_{\rm QCD}$, the collinear factorization at twist-3 level~\cite{Qiu:1991pp,Qiu:1991wg} plays the central role and SSAs are generated by the correlations of multi-parton densities in the nucleon, the so-called collinear twist-3 functions~\cite{Efremov:1981sh, Efremov:1984ip,Qiu:1991pp, Qiu:1991wg}. It was theoretically proven and demonstrated phenomenologically that in the intermediate region, $Q_2 \gtrsim Q_1\gg \Lambda_{\rm QCD}$, the two formalisms are related~\cite{Ji:2006ub,Ji:2006br,Koike:2007dg,Yuan:2009dw,Zhou:2009jm,Cammarota:2020qcw}. TMDs can be expressed in terms of collinear and twist-3 functions via Operator Product Expansion~\cite{Aybat:2011ge,Kanazawa:2015ajw,Scimemi:2019gge}. The integral relations~\cite{Boer:2003cm,Gamberg:2017jha,Qiu:2020oqr} based on operator definitions also show the close relation between TMD and twist-3 distributions.

Inclusive jet production in $pp$ collisions is one example of single-scale processes that can be described within the twist-3 approach, see for instance Ref.~\cite{Gamberg:2013kla}. Nevertheless, other approaches, such as the so-called generalized parton model (GPM)~\cite{DAlesio:2004eso,Anselmino:2005sh,DAlesio:2007bjf}, where a factorized formulation in terms of TMDs is assumed as the starting point, have been successfully applied in the analysis of SSAs for inclusive particle production in $pp$ collisions, as well as in inclusive jet production~\cite{Anselmino:2013rya}. The GPM is indeed a very powerful method to study processes where factorization is not established and can, hopefully, shed light on factorization breaking effects. 

In the last decade, a color gauge invariant formulation of the GPM, named CGI-GPM, has been proposed~\cite{Gamberg:2010tj} and then extensively developed in Refs.~\cite{DAlesio:2011kkm, DAlesio:2017rzj,DAlesio:2018rnv}. Its main feature is the inclusion of initial- (ISI) and final- (FSI) state interactions within a one-gluon exchange approximation. As a result, the Sivers function becomes non-universal, and its calculable process dependence can be absorbed into the partonic cross sections. Hence, in the evaluation of physical observables, one can still use the quark Sivers function obtained from SIDIS fits, but now convoluted with modified partonic cross sections, such that the expected sign change from SIDIS to DY is restored. Moreover, this modified GPM formalism has a very close connection~\cite{Gamberg:2010tj} with the collinear twist-3 approach. In all these respects, it is extremely interesting to explore its potentiality and its implications. 

In the spirit of testing the compatibility of the extraction of the Sivers function, as obtained by  best-fitting the SIDIS azimuthal asymmetries, we analyze the recent SSA data for inclusive jet production in $pp$ collisions from the STAR Collaboration at RHIC~\cite{Adam:2020edg}, within the GPM and the CGI-GPM approaches. These data cover a wide region of $x$, expanding the range explored in SIDIS measurements, and will allow us to improve and extend our current knowledge on the quark Sivers function.

As a global fit would at present require prohibitive machine power, here we employ an equivalent, but less numerically costly procedure, the so-called reweighting method~\cite{Giele:1998gw,Ball:2010gb,Sato:2013ika,Sato:2016wqj,Lin:2017stx,DAlesio:2020vtw} within Bayesian statistics. Reweighting allows us to properly include the information from the new set of data in the phenomenological analysis, estimate their impact on the extraction of the quark Sivers distributions,  and determine via Bayesian statistics the fitted parameters and their errors.  
At the same time, this will also allow us to test the relevance of the expected process dependence of the Sivers function. 

The paper is organized as follows. In Section~\ref{sec:GPM-and-CGI} we recall the formalism applied to compute the SIDIS azimuthal Sivers asymmetry and the corresponding single spin asymmetries for inclusive jet production in $pp$ collisions. In Section~\ref{sec:reweighting}, we discuss the reweighting procedure and extend it to the case of asymmetric errors. In Section~\ref{sec:results} we apply reweighting to obtain the statistical impact of the new STAR data sets to the extraction of the Sivers function. Finally, in Section~\ref{sec:conclusions}, we draw our conclusions. 

\section{\label{sec:GPM-and-CGI} The formalism}

Here we briefly recall the main aspects of the theoretical formalism employed in this study. All details can be found in the papers quoted below.

The expression for the azimuthal Sivers asymmetry in SIDIS, $\ell p^\uparrow  \to \ell' h X$,  is given by~\cite{Bacchetta:2006tn,Anselmino:2011ch}
\begin{equation}
\label{eq:siv-asy}
A_{UT}^{\sin(\phi_h-\phi_S)}  \equiv 
\frac{F_{UT}^{\sin(\phi_h-\phi_S)}}{F_{UU,T}} \,,
\end{equation}
where $F_{UU,T}= \mathcal{C}[f_1^q D_1^q]$ is the  TMD unpolarized structure function, and $F_{UT}^{\sin(\phi_h-\phi_S)}= \mathcal{C}[f_{1T}^{\perp q} D_1^q]$ is the azimuthal modulation triggered by the correlation between the nucleon spin and the quark intrinsic transverse momentum. This effect is embodied in the Sivers function~\cite{Bacchetta:2004jz} 
\begin{eqnarray}
\Delta\hat{f}_{q/p^\uparrow}(x,\bm{k}_\perp) & = &  \Delta^Nf_{q/p^\uparrow}(x,k_\perp) \sin(\phi_S-\varphi) \nonumber \\
&=& \!\!\!\! -\frac{2k_{\perp}}{M_p} f^{\perp q}_{1 T}(x, k_{\perp})\sin(\phi_S-\varphi) \,,
\end{eqnarray} 
which appears in the number density of unpolarized quarks, $q$, with intrinsic transverse momentum $\bm{k}_\perp= k_\perp(\cos\varphi, \sin\varphi)$, inside a transversely polarized proton $p^\uparrow$, with polarization vector $\bm{S}_T=S_T(\cos\phi_S, \sin\phi_S)$ and mass $M_p$, moving along the $z$ direction. 
The dependence of structure functions and TMDs on the hard scale $Q^2$ is omitted for brevity.

For the inclusive jet production in $pp$ collisions, the SSA is defined as \begin{equation}\label{eq:AN}
A_N \equiv \frac{d\sigma^\uparrow-d\sigma^\downarrow}{d\sigma^\uparrow+d\sigma^\downarrow} \equiv\, \frac{d\Delta\sigma}{ 2 d\sigma}\,,
\end{equation}
where $d\sigma^{\uparrow(\downarrow)}$ denotes the single-polarized cross section, in which one of the protons is polarized along the transverse direction $\uparrow$($\downarrow$) with respect to the production plane.
For this process, within the GPM as well as the CGI-GPM approach, only the Sivers effect, from quarks and gluons, can be at work.\footnote{A second possible TMD effect, coming from the convolution of the transversity distribution and the Boer-Mulders function, turns out to be negligible.}
The  gluon  contribution is negligible in the  region  of moderate and forward rapidities, as well as at small $x_F$ ($x_F= 2P_{jL}/\sqrt s$, with $P_{jL}$ the longitudinal jet momentum) as shown, for both approaches, in Refs.~\cite{DAlesio:2015fwo, DAlesio:2018rnv}. The gluon Sivers effect can therefore be safely ignored in this study.

Within the framework of the CGI-GPM, the numerator of the asymmetry is given by~\cite{Gamberg:2010tj}
\begin{equation}
\label{eq:sivgen}
\begin{aligned}
 d\Delta&\sigma^{\rm CGI-GPM}\, \equiv  \,  \frac{E_j \, d\sigma^\uparrow}{d^3\bm{P}_j} -
\frac{E_j \, d\sigma^\downarrow}{d^3\bm{P}_j} \\
& = \frac{2\alpha_s^2}{s}\sum_{a,b,c,d} \int \frac{dx_a \, dx_b}{ x_a \, x_b} \; d^2\bm{k}_{\perp a} \, d^2\bm{k}_{\perp b}\\
& \times \left ( -\frac{k_{\perp a}}{M_p} \right  ) f^{\perp a}_{1 T}(x_a, k_{\perp a})\cos\varphi_a\\
 &\times  f_{b/p}(x_b, k_{\perp b})\,H^{\rm Inc}_{ab \to cd}
\> \delta(\hat s + \hat t + \hat u) \>,
\end{aligned}
\end{equation}
where $\alpha_s$ is the strong coupling constant, $s$ is the center-of-mass (c.m.) energy, $\bm{P}_{j}$ is the jet momentum, and  $\hat s$, $\hat t $, $\hat u$ are the usual Mandelstam variables for the partonic subprocess $ab\to cd$. Furthermore, $f_{b/p}(x_b, k_{\perp b})$ is the TMD distribution for an unpolarized parton $b$ inside the unpolarized proton. Notice that in a leading-order (LO) approach the jet is identified with the final parton $c$. 
Moreover, $H^{\rm Inc}_{ab \to cd}$ are the perturbatively calculable hard scattering functions. 
In particular, the $H^{\rm Inc}_{ab \to cd}$ functions where $a$ is a quark or an antiquark can be found in Ref.~\cite{Gamberg:2010tj}.
The GPM results can be obtained from \cref{eq:sivgen} by simply replacing $H^{\rm Inc}_{ab \to cd}$ with the standard unpolarized partonic cross sections, $H^U_{ab \to cd}$. 

Finally, the unpolarized cross section, $d\sigma$, at denominator in~\cref{eq:AN}, is
\begin{equation}
\label{eq:unp}
\begin{aligned}
 d\sigma\,
& = \frac{\alpha_s^2}{s}\sum_{a,b,c,d} \int \frac{dx_a \, dx_b}{ x_a \, x_b} \; d^2\bm{k}_{\perp a} \, d^2\bm{k}_{\perp b}\\
& \times f_{a/p}(x_a, k_{\perp a}) 
   f_{b/p}(x_b, k_{\perp b})\,H^U_{ab \to cd}
\> \delta(\hat s + \hat t + \hat u) \>.
\end{aligned}
\end{equation}
In the computation of the jet SSA we will adopt the jet transverse momentum $P_{jT}$ as the factorization scale. 

\section{\label{sec:reweighting} The reweigthing method for TMDs}

We now briefly illustrate the reweighting method that we will apply in our study. This is a well established technique in the context of collinear PDF extractions. Since the seminal work of Giele and Keller~\cite{Giele:1998gw}, the method has been used in both the Bayesian framework~\cite{Ball:2010gb,Ball:2011gg,Sato:2013ika,Sato:2016wqj,Lin:2017stx,DAlesio:2020vtw} and in the Hessian approach~\cite{Paukkunen:2014zia,Schmidt:2018hvu,Eskola:2019dui}. 
The reweighting procedure allows to assess the impact of new data sets on extractions of distributions describing hadron structure, avoiding a new global fit. It also indicates whether these additional data are consistent with the data sets used for the original extraction. 

Let us consider a model for TMDs depending on a $n$-dimensional set of parameters $\bfa = \{a_1, \ldots, a_n\}$. Traditionally, a $\chi^2$ minimization procedure is employed in order to estimate the values of parameters that describe the experimental data. Let us suppose that a set of data $\bfy = y_1, \ldots, y_{N_{\rm dat}}$ (with an associated covariance matrix $C$) is measured. Then, the $\chi^2$ is defined as~\footnote{If only uncorrelated uncertainties, $\sigma_i$, are given, the new $\chi^2$ reduces simply to $\chi^2[\bfa,\bfy]= \sum\limits_{i = 1}^{N_{\rm dat}} \dfrac{(y_i[\bfa] - y_i)^2}{\sigma_i^2}$.} 
\begin{equation}
    \chi^2[\bfa, \bfy] = \sum_{i,j = 1}^{N_{\rm dat}} (y_i[\bfa] - y_i)\, C_{ij}^{-1}(y_j[\bfa] - y_j)\,,
\end{equation}
where we have indicated by $y_i[\bfa]$ the values computed using the theoretical model.
The best-fit set $\bfa_0$, determined through the $\chi^2$ minimization procedure, will have a corresponding minimum value $\chi^2_0[\bfa_0]$.
The uncertainty on the extracted TMDs can either be calculated in terms of Hessian eigensets or generated by applying Monte Carlo (MC) procedures. In an ideal case, both methods yield the correct results and can be used in phenomenology. However, the former relies on the Gaussianity of the underlying distributions and does not necessarily account for the tails of distributions or for the potential presence of multiple solutions to the minimum of $\chi^2$ in the parameter space.
In order to circumvent these complications, we will use a reweighting procedure based on the  Bayesian inference. This will allow us to exploit the well known advantages of Bayesian inference, \ie the ability to construct probability distributions for the parameters, and study the influence of the new data on the prior knowledge.

Let us assume that some prior knowledge, say theoretical, exists on the parameters and that they are described by probability density functions $\pi(\bfa)$, \ie the prior distributions. If no prior knowledge exists, $\pi(\bfa)$ are flat distributions.

We now generate the parameter sets, $\bfa_k$ ($k = 1, \ldots, N_{\rm set}$), with corresponding $\chi^2_{k} \in [\chi^2_0, \chi^2_0 + \Delta\chi^2]$, where $\Delta\chi^2$ is the desired tolerance corresponding to $n$ parameters and at a certain confidence level (CL). For this purpose, we adopt a Markov Chain Monte Carlo procedure and produce $N_{\rm set}$ parameter sets, employing a Metropolis-Hastings algorithm with an auto-regressive generating density~\cite{10.2307/2684568}.
Such approach allows for an efficient exploration and reconstruction of the parameter space, starting from the information contained in the error matrix obtained from the minimization procedure.
The $\chi^2_k$ for every parameter set $\bfa_k$ is calculated as:
\begin{equation}
    \chi^2_k[\bfa_k, \bfy] = \sum_{i,j = 1}^{N_{\rm dat}} (y_i[\bfa_k] - y_i)\, C_{ij}^{-1}(y_j[\bfa_k] - y_j)\,.
    \label{eq:chi2k}
\end{equation}
Using Bayes theorem, one can then calculate the posterior density given the data set as
\begin{equation}
   \cal{P}(\bfa| \bfy )=\frac{{\cal L} (\bfy | \bfa)\, \pi(\bfa)}{\mathit{Z}}\; ,
   \label{eq:posterior}
\end{equation}
where ${\cal L} (\bfy| \bfa)$ is the likelihood, and $Z = {\cal P}(\bfy)$ is the evidence, that ensures a normalized posterior density. $\cal{P}(\bfa|\bfy)$ will therefore incorporate the impact of the data on our knowledge of TMDs. 

Different choices for the likelihood have been discussed in the literature. Following Refs.~\cite{Giele:1998gw,Sato:2013ika,Sato:2016wqj,Lin:2017stx,DAlesio:2020vtw}, here we adopt the likelihood definition as obtained by taking ${\cal L} (\bfy| \bfa)\,d\bfy$ as the probability to find the new data confined in a differential volume $d\bfy$ around $\bfy$. This results in defining the weights $\wk$ as follows: 
\begin{equation}\label{eq:weights}
 \wk (\chi^2_k) = \frac{\rm{exp}\left\{-\frac 12\,\chi^2_\itk[\bfa_\itk,\bfy]\right\}}{\sum\limits_i \mathit{w}_{\mathit i}}\,.
\end{equation}
Such weights are normalized to 1, and can be used to calculate, for any given observable $\cal{O}$, its expectation value $\rm{E}[\cal{O}]$ and variance $\rm{V}[\cal{O}]$ as
\begin{equation}\label{eq:E}
 \rm{E}[\mathcal{O}] \simeq \sum_{\mathit k} \wk\,\mathcal{O}(\bfa_{\mathit{k}})\,,
\end{equation}
\begin{equation}\label{eq:V}
 \rm{V}[\mathcal{O}] \simeq \sum_{\itk}\wk\left(\mathcal{O}(\bfa_\itk) - {\rm{E}}[\mathcal{O}]\right) ^2.
\end{equation}
Note that~\cref{eq:E,eq:V} are related to a symmetric error calculation. It is well known that when a parameter is symmetrically distributed (\eg according to a Gaussian distribution) the mean value will coincide with the median value and the uncertainty at a certain confidence level, for instance at $68\%$ (1$\sigma$) or $95\%$ (2$\sigma$) will be symmetric. Still, this does not ensure that any distribution for any observable $\mathcal{O}(\bfa_k)$, depending on Gaussian distributed parameters $\bfa_k$, is itself Gaussian.

In fact, asymmetric distributions may arise. Thus, mean and median for $\mathcal{O}(\bfa_k)$ would, in general, be different: the more the distribution is asymmetric, the more the uncertainty on the observable is not properly described by a symmetrized error, as discussed for example in Ref.~\cite{Possolo_2019}. To overcome this potential issue, we extend the reweighting method, providing asymmetric uncertainties. Such errors are calculated using the \texttt{rv\_discrete} function of the \texttt{SciPy} Python library~\cite{2020SciPy-NMeth}. This function is able to build a discrete weighted distribution, providing automatically the mean value, the median and the asymmetric uncertainty interval at a specific confidence level. The interval is estimated considering equal areas around the median, and the endpoints of these areas are given automatically by \texttt{rv\_discrete}. For our analysis we will adopt the median as central value, and the uncertainty will be provided at a $2\sigma$ CL.
We will perform this procedure with the data and parametrizations considered in Ref.~\cite{Boglione:2018dqd} and explicitly demonstrate the equivalence of the reweighting procedure and the global fit.

The Bayesian inference is easily generalized to evaluate the impact of the new data $\bfy_{\rm new}$, in our case the data from STAR Collaboration at RHIC~\cite{Adam:2020edg}.
As a matter of fact, the new evidence, \ie the new data, will change the weights $\wk \to \wk^{\rm new} \equiv \wk (\chi^2_{k}+\chi^2_{\rm{new}, k})$ for each parameter set $\bfa_k$, where $\chi^2_{\rm{new}, k}=\chi^2_{k}[\bfa_k, \bfy_{\rm new}]$, Eq.~\eqref{eq:chi2k}.
Therefore, the probability distribution for every parameter, as well as any other observable that depends on the TMDs, is expected to change. They can be calculated either with initial priors $\pi(\bfa)$ and weights  $\wk (\chi^2_{k}+\chi^2_{\rm{new}, k})$  or, equivalently, with posteriors $\cal{P}(\bfa| \bfy )$ after the global fit as the priors of reweighting and weights  $\wk(\chi^2_{\rm{new}, k})$. Notice that, as expected, weights are multiplicative, $\wk(\chi^2_{k}+\chi^2_{\rm{new}, k}) \propto  \wk(\chi^2_{k})\wk(\chi^2_{\rm{new}, k})$, as $\chi^2$ is additive.
The resulting posteriors will contain the impact of the new data and the new values of the parameters, and the observables can be evaluated with Eqs.~\eqref{eq:E}, \eqref{eq:V} or with \texttt{rv\_discrete} function. As already mentioned, we will adopt the latter in our analysis.

In the following, we will apply the reweighting technique for the first time to a TMD function, and in particular to the quark Sivers functions. 

\section{\label{sec:results} Results}

We will now present and discuss our results on the reweighting of the quark Sivers function. Before showing the outcome of the reweighting procedure using jet $A_N$ data from STAR, let us briefly illustrate the model parametrization chosen to describe the Sivers function. 

Here we adopt the quark Sivers function extracted from SIDIS data in Ref.~\cite{Boglione:2018dqd}. More precisely, we use the so called ``reference fit", which results in a minimum $\chi^2_{\rm dof} = 0.99$ for $N_{\rm dat}^{\rm SIDIS} = 220$ data points (see Table 2 of Ref.~\cite{Boglione:2018dqd}). 
Notice that this choice is motivated by the simplicity of this fit, which makes it particularly  suitable for the purposes of this work. The relatively small number of parameters will allow us to highlight the effects of reweighting the Sivers function. Another important aspect is that this fit was performed paying special attention to the amount of information one can actually infer from data, reducing the assumptions which could bias the extraction. In this respect, we will be able to show and quantify the impact of a new set of data on our knowledge of the Sivers function using Bayesian inference.

The ``reference fit" consists in a factorized $x$ and $k_\perp$ dependence (the latter being Gaussian-like and flavor independent) for the up- and down-quark Sivers function. Within this parametrization, the Sivers function reads
\begin{equation}\label{eq:Sivers-parametrization}
    \Delta^N\!f_{q/p^\uparrow}(x,k_\perp) = \frac{4 M_p k_\perp}{\langle k_\perp^2\rangle_S} \Delta^N\!f_{q/p^\uparrow}^{(1)}(x) \frac{e^{-k_\perp^2/\langle k_\perp^2\rangle_S}}{\pi \langle k_\perp^2\rangle_S}\,,
\end{equation}
where $q = u, d$, and where $\Delta^N\!f_{q/p^\uparrow}^{(1)}(x)$ is the Sivers first $k_\perp$-moment:
\begin{equation}\label{eq:f1Tp-first-mom}
\begin{aligned}
    \Delta^N\!f_{q/p^\uparrow}^{(1)}(x)  & = \int d^2 \bfk_\perp \frac{k_\perp}{4 M_p} \Delta^N\!f_{q/p^\uparrow}(x,k_\perp) \equiv - f_{1T}^{\perp (1) q}(x) \\
    &= N_q\,(1-x)^{\beta_q}\,.
\end{aligned}    
\end{equation}
Thus, the model depends on a total of five parameters: $N_u, N_d, \beta_u, \beta_d,$ and $\langle k_\perp^2\rangle_S$. 

In the computation of any asymmetry, special care should be taken in the calculation of the corresponding unpolarized cross section. Here we follow the same approach adopted in Ref.~\cite{Boglione:2018dqd}, and compute the unpolarized SIDIS cross section, which appears at denominator in Eq.~(\ref{eq:siv-asy}), by applying the $k_\perp$-widths extracted in the multiplicity analysis of Ref.~\cite{Anselmino:2013lza}. For the jet single spin asymmetry, we adopt the corresponding $k_\perp$-width resulting from the fit of HERMES SIDIS data.

For the collinear parton densities, we use the CTEQ6L1 set of PDFs~\cite{Stump:2003yu} and the DSS set of fragmentation functions~\cite{deFlorian:2007aj}. Finally, for this fit, we generate $N_{\rm set} = 2 \cdot 10^5$ parameter sets, adopting the corresponding $\Delta\chi^2$ at $2\sigma$ CL for $N = 5$ parameters, \ie $\Delta\chi^2 = 11.31$, as tolerance.

\subsection{\label{sec:SIDIS-jet-reweighting} Impact of jet data on the quark Sivers extraction}

We now proceed by illustrating our final results. Very recently, the STAR Collaboration at RHIC has published new measurements of the single-spin asymmetries in inclusive jet production from polarized proton-proton collisions, at two different c.m.~energies: $\sqrt{s} = 200\GeV$ and $\sqrt{s} = 500\GeV$~\cite{Adam:2020edg}. This new data set, amounting to $N_{\rm dat}^{\rm{jet}} = 18$ points, covers a wide range in $x_F$ ($0.1 \lesssim x_F \lesssim 0.6$), and is useful to further constrain the quark Sivers function in the large-$x$ region, where information from SIDIS data is either scarce or even absent. 
\begin{figure}[t]
\centering
\includegraphics[width=\columnwidth]{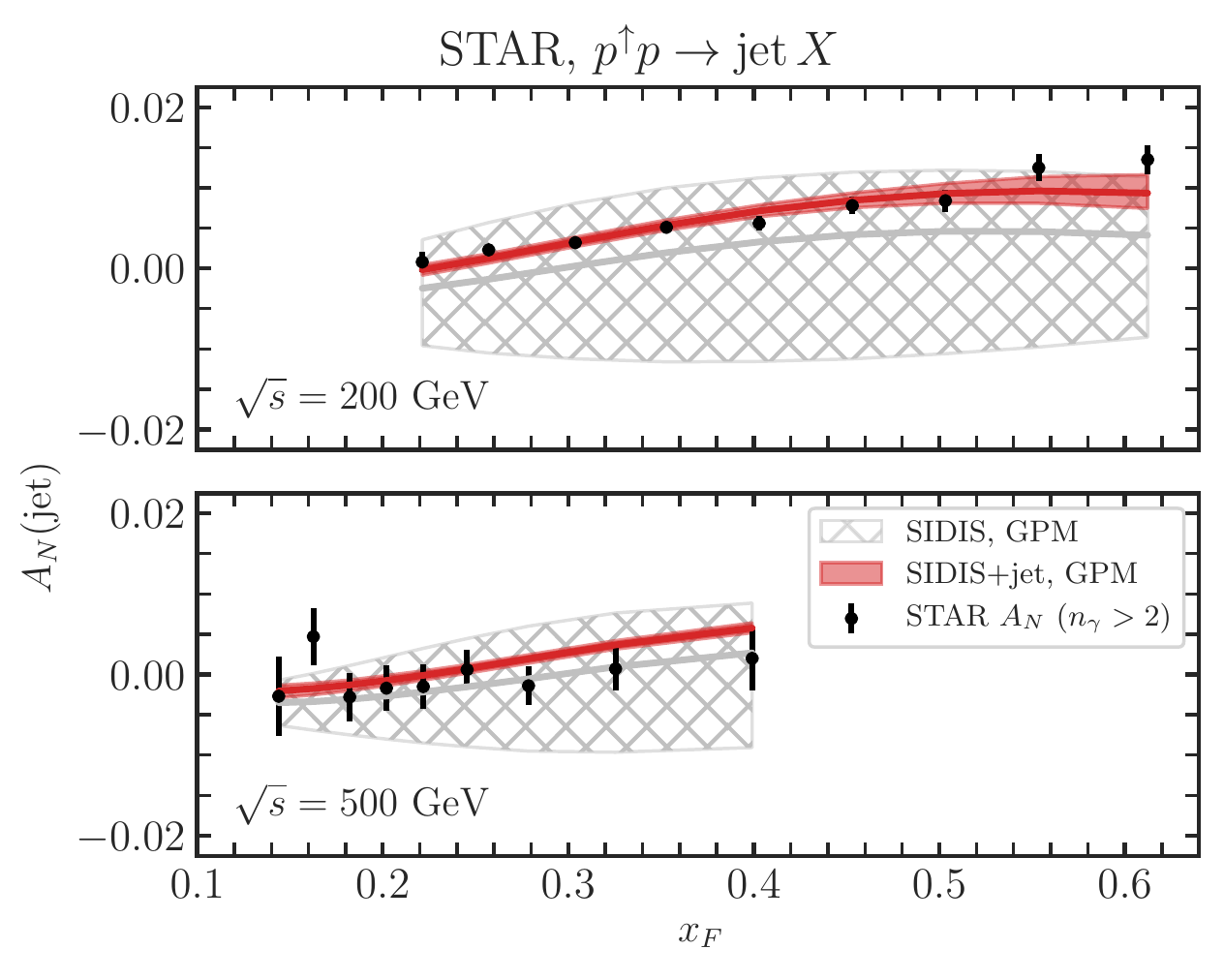}
\caption{Results for the reweighting procedure from SIDIS and $A_N$ jet data in the GPM formalism, compared with STAR measurements~\cite{Adam:2020edg} at $\sqrt{s} = 200\,\text{GeV}$ (upper panel) and $\sqrt{s} = 500\,\text{GeV}$ (lower panel). Uncertainty bands are at $2\sigma$ CL. The results before (hatched grey bands) and after (solid red bands) reweighting are shown.}
\label{fig:reweigthing-SIDIS+jet-GPM}
\end{figure}
STAR measurements refer to electromagnetic jet production from polarized $pp$ scattering. Two sets of data are collected at each energy: one fully inclusive and one imposing a cut on the photon multiplicity ($n_\gamma>2$). The latter is the most suitable for our analysis since it is not contaminated by single-photon and isolated $\pi^0$ production. Nevertheless, for completeness, we have also considered this data set and we will comment on the corresponding results.

In the following, apart from showing the impact of the $A_N$ jet data on the extraction of the Sivers function, we will also address whether any signal pointing towards a process dependence of the Sivers function itself can be observed.

As far as SIDIS is concerned, for the extraction of the Sivers function we will refer to Ref.~\cite{Boglione:2018dqd}, applying a LO approximation. Indeed, extractions of the Sivers function at higher perturbative orders exist, as in Refs.~\cite{Bacchetta:2020gko,Echevarria:2020hpy,Bury:2020vhj}, but all the extracted Sivers functions are in good agreement, confirming the findings of Ref.~\cite{Boglione:2018dqd} on the weak dependence of the asymmetries on the scale. We plan to perform similar TMD phenomenological analyses and reweighting of the Sivers functions to higher perturbative orders in the future.

In what follows, the predictions based on the Sivers functions as extracted by fitting only the SIDIS data are dubbed as ``SIDIS", while the asymmetries computed after the reweighting procedure, by using the jet data as the new evidence as described in \cref{sec:reweighting}, are indicated by a ``SIDIS+jet" label. The central value and the asymmetric uncertainty bands at $2\sigma$ CL are calculated using the procedure explained in~\cref{sec:reweighting}, and adopting the corresponding weights for the ``SIDIS" and the ``SIDIS+jet" (GPM and CGI-GPM) cases.

\begin{figure}[t]
\centering
\includegraphics[width=\columnwidth]{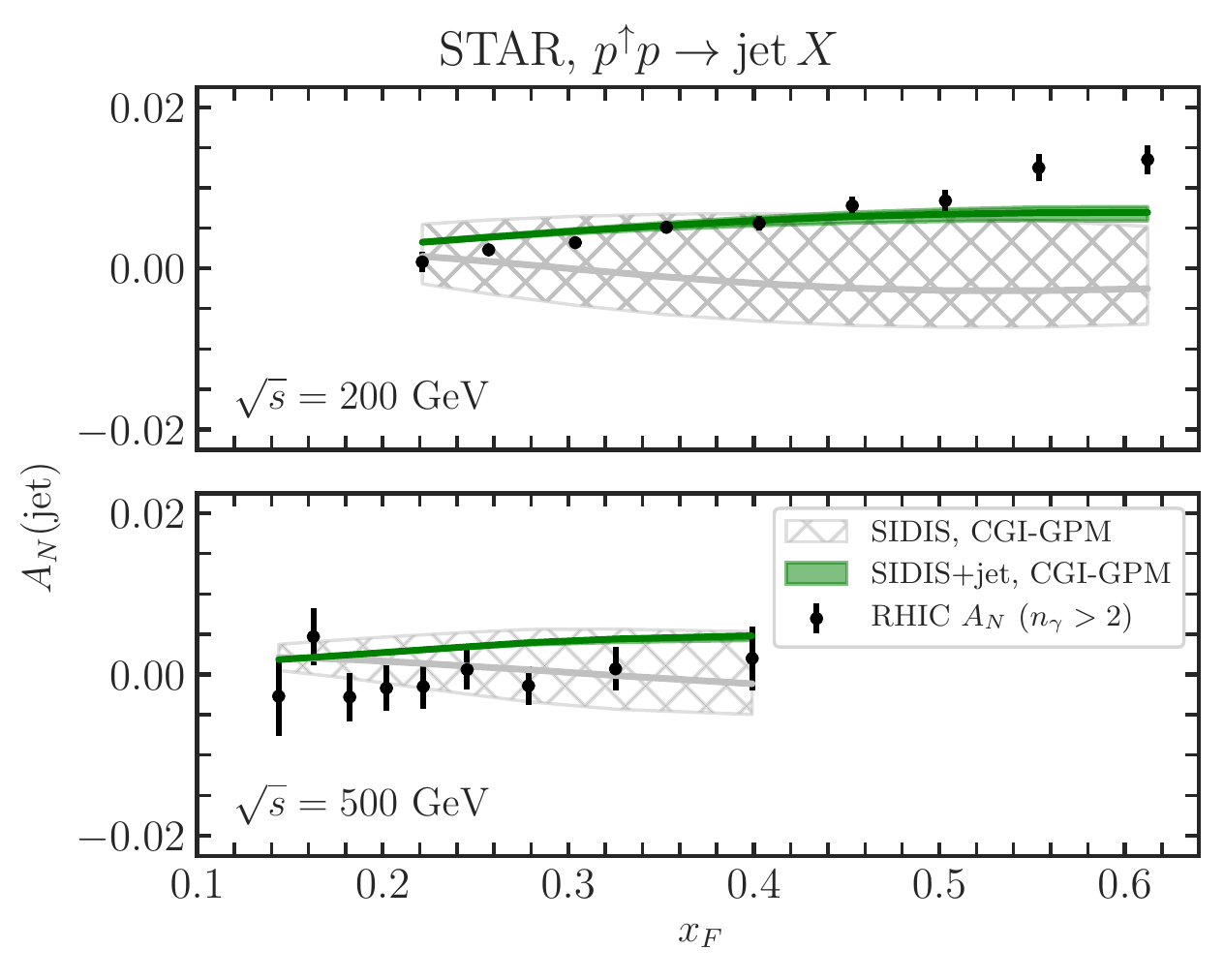}
\caption{Results for the reweighting procedure from SIDIS and $A_N$ jet data in the CGI-GPM formalism, compared with STAR measurements~\cite{Adam:2020edg} at $\sqrt{s} = 200\,\text{GeV}$ (upper panel) and $\sqrt{s} = 500\,\text{GeV}$ (lower panel). Uncertainty bands are at $2\sigma$ CL. The results before (hatched grey bands) and after (solid green bands) reweighting are shown.}
\label{fig:reweigthing-SIDIS+jet-CGI}
\end{figure}

\cref{fig:reweigthing-SIDIS+jet-GPM,fig:reweigthing-SIDIS+jet-CGI} show the results of the reweighting procedure for the Sivers function in the GPM and the CGI-GPM, respectively. On the upper (lower) panels, the comparison with data at $\sqrt{s} = 200\,(500)\GeV$ is shown.

\begin{table*}[htb]
\centering
\begin{tabular}{c c c c c c c}
\toprule
 ~& $N_u$ & $N_d$ & $\beta_u$ & $\beta_d$ & $\langle k_\perp^2\rangle_S$ & $\chi^2_{\rm dof}$ \\
\midrule
SIDIS & $0.40^{+0.05}_{-0.04}$ & $-0.65^{+0.13}_{-0.15}$ & $5.52^{+0.93}_{-0.83}$ & $6.77^{+2.29}_{-1.85}$ & $0.30^{+0.08}_{-0.08}$ & $1.01^{+0.03}_{-0.02}$\\
\midrule
SIDIS+jet, GPM & $0.36^{+0.04}_{-0.03}$ & $-0.55^{+0.07}_{-0.10}$ & $4.98^{+0.34}_{-0.30}$ & $6.45^{+0.63}_{-0.52}$ & $0.28^{+0.07}_{-0.07}$ & $1.05^{+0.03}_{-0.01}$\\
\midrule
SIDIS+jet, CGI-GPM & $0.35^{+0.02}_{-0.01}$ & $-0.43^{+0.01}_{-0.02}$ & $4.79^{+0.28}_{-0.19}$ & $4.48^{+0.17}_{-0.13}$ & $0.26^{+0.03}_{-0.02}$ & $1.25^{+0.04}_{-0.01}$\\
\bottomrule
\end{tabular}
 \caption{Summary of the results of the reweighting procedure for the fitted parameters. The quoted asymmetric uncertainties are at $2\sigma$ CL.}
\label{tab:results}
\end{table*}

The results in~\cref{fig:reweigthing-SIDIS+jet-GPM,fig:reweigthing-SIDIS+jet-CGI} clearly show the effect of the reweighting procedure. As we will see later, the jet SSA data measured by the STAR Collaboration allow to drastically reduce the uncertainty on the quark Sivers functions (especially for $d$ quarks) and, in turn, on the corresponding estimates for the single-spin asymmetries in $p^\uparrow p \to {\rm jet}\,X$. 
It is important to emphasize the role of the STAR data, that extend up to $x_F \simeq 0.6$ with remarkably small uncertainties, offering valuable information on the large $x$-region which in SIDIS remains largely uncovered.
The reweighting procedure clearly indicates that jet data offer a crucial complementary information to SIDIS data. This analysis also points in favor of a good compatibility between the two data sets.

\begin{figure}[t]
\centering
\includegraphics[width=\columnwidth]{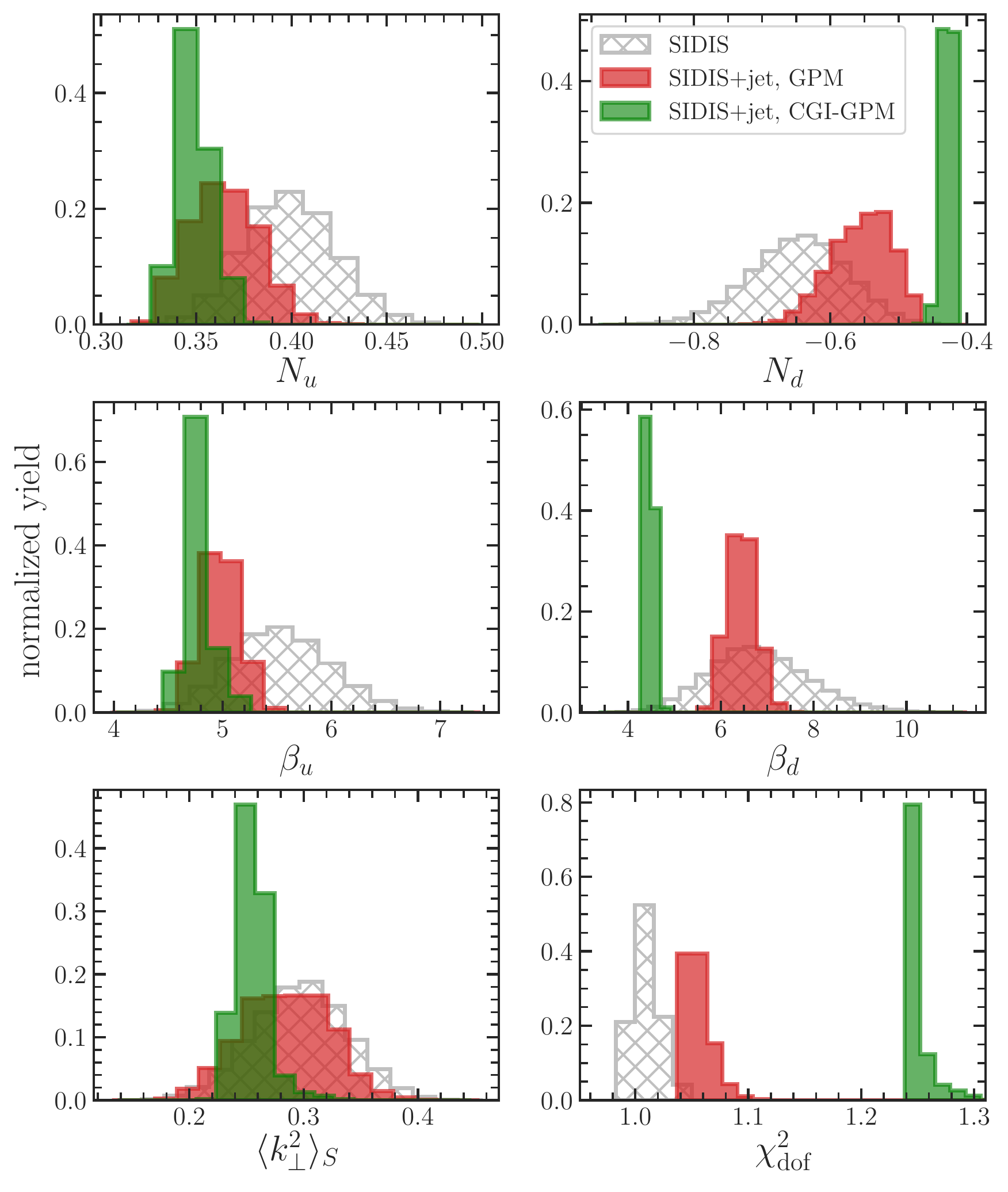}
\caption{Parameters and $\chi^2_{\rm dof}$ probability densities. Hatched histograms refer to the priors coming from SIDIS data only; the red (green) ones are the posterior densities, reweighted using jet data, in the GPM (CGI-GPM) formalism.}
\label{fig:pars-chi2ndof-distributions}
\end{figure}

To better interpret the results shown in~\cref{fig:reweigthing-SIDIS+jet-GPM,fig:reweigthing-SIDIS+jet-CGI}, in~\cref{fig:pars-chi2ndof-distributions} we present the probability densities before and after reweighting for the five parameters of the ``reference fit'' and the $\chi^2_{\rm dof}$. 
The corresponding values and uncertainties, computed according to the procedure described at the end of~\cref{sec:reweighting}, are gathered in~\cref{tab:results}. Notice that the results for the ``SIDIS" case are fully compatible with the findings of Ref.~\cite{Boglione:2018dqd}.
Some comments are in order:
\begin{enumerate}[(i)]
\item The flavor independent Sivers Gaussian width (lower left panel in~\cref{fig:pars-chi2ndof-distributions}) does not vary significantly. This signals a mild role of  TMD-evolution effects in the available jet SSA data. It is indeed worth to notice that at large $x_F$ we probe large $P_{jT}$ values (up to $4\div 5\GeV$ depending on the data set). Since $P_{jT}$ represents the factorization scale adopted for this observable, we reach $Q^2$ scales significantly larger than those probed in SIDIS. However, asymmetries are ratios of cross sections where evolution and higher order effects tend to cancel out~\cite{Kang:2017btw}. Although our parametrization does not have the complete features of TMD evolution, results of Refs.~\cite{Boglione:2018dqd} are compatible with full TMD evolution at higher logarithmic accuracy~\cite{Bacchetta:2020gko,Echevarria:2020hpy,Bury:2020vhj}.
\item The $\beta_q\,(q = u,d)$ parameters (mid panels in~\cref{fig:pars-chi2ndof-distributions}) that govern the large-$x$ behavior of the Sivers function change, but in a different way when applying the GPM or the CGI-GPM formalisms: this is due to the fact that in the CGI-GPM approach color factors change the role of the $u$- and $d$-quark Sivers contributions. In particular, for the dominant channels in the forward rapidity region, like $qg\to qg$, $H^{\rm inc}$ in the CGI-GPM approach presents, roughly, a change of sign w.r.t.~$H^U$ in the GPM (which are all positive). This implies that while the slightly positive $A_N$ at large $x_F$ in the GPM is driven by the positive sign of the up-quark Sivers function, in the CGI-GPM is given by the negative down-quark Sivers function. 
\item Concerning the normalization parameters (upper panels in~\cref{fig:pars-chi2ndof-distributions}) we see that while $N_u$ changes slightly, $N_d$ turns out to be smaller in size in the CGI-GPM approach. On the other hand, as mentioned above, the corresponding Sivers function for $d$-quarks is less suppressed in the large-$x$ region.
\item The $\chi^2_{\rm dof}$ after the reweighting, calculated for $N_{\rm dat}^{\rm{SIDIS+jet}} = 238$ points, is different for the GPM and the CGI-GPM cases, slightly favoring the former approach. \end{enumerate}

\begin{figure}[t]
\centering
\subfloat[]{\includegraphics[width=\columnwidth]{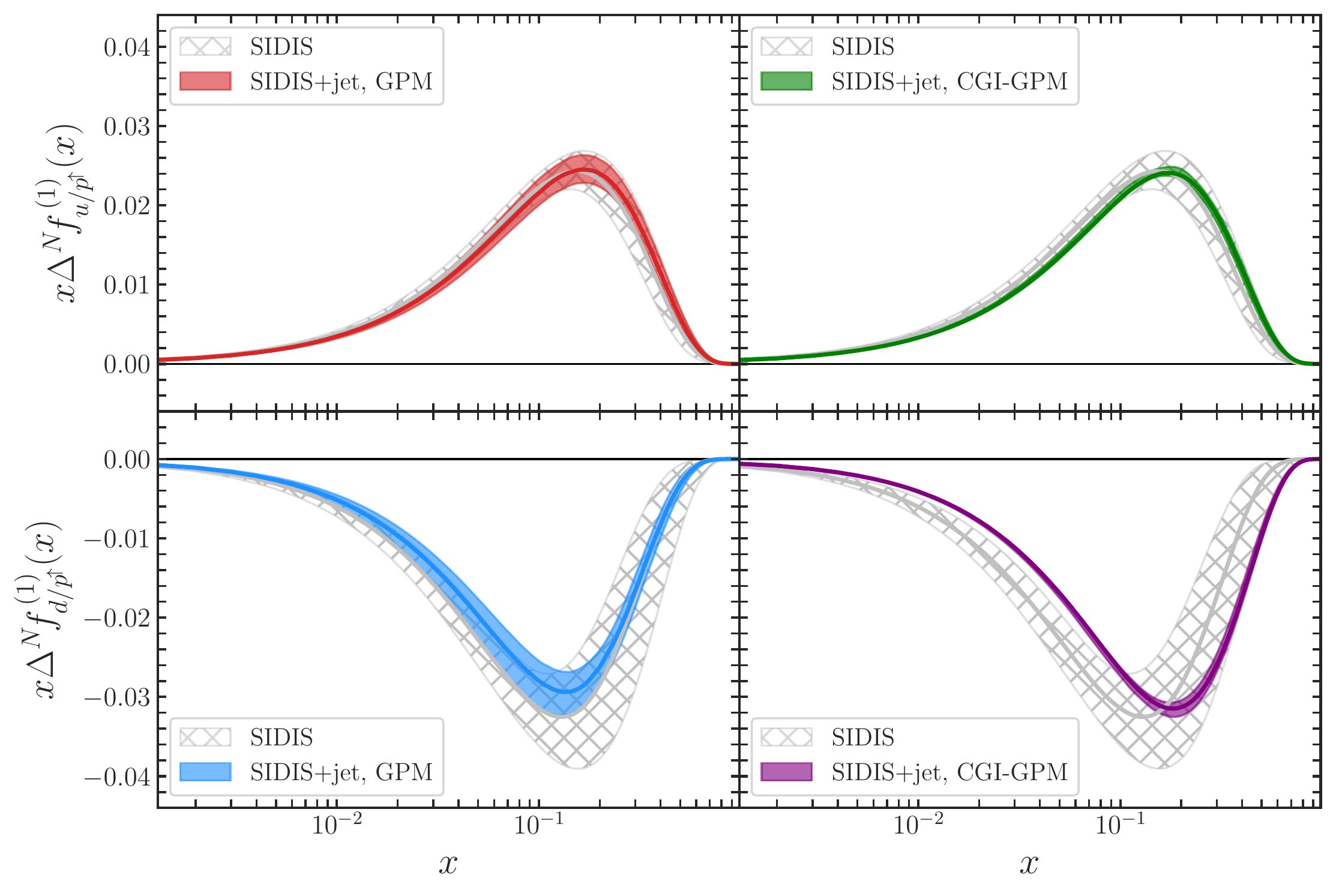}}\label{fig:siv-comparison}
\subfloat[]{\hspace{4mm}\includegraphics[width=.955\columnwidth]{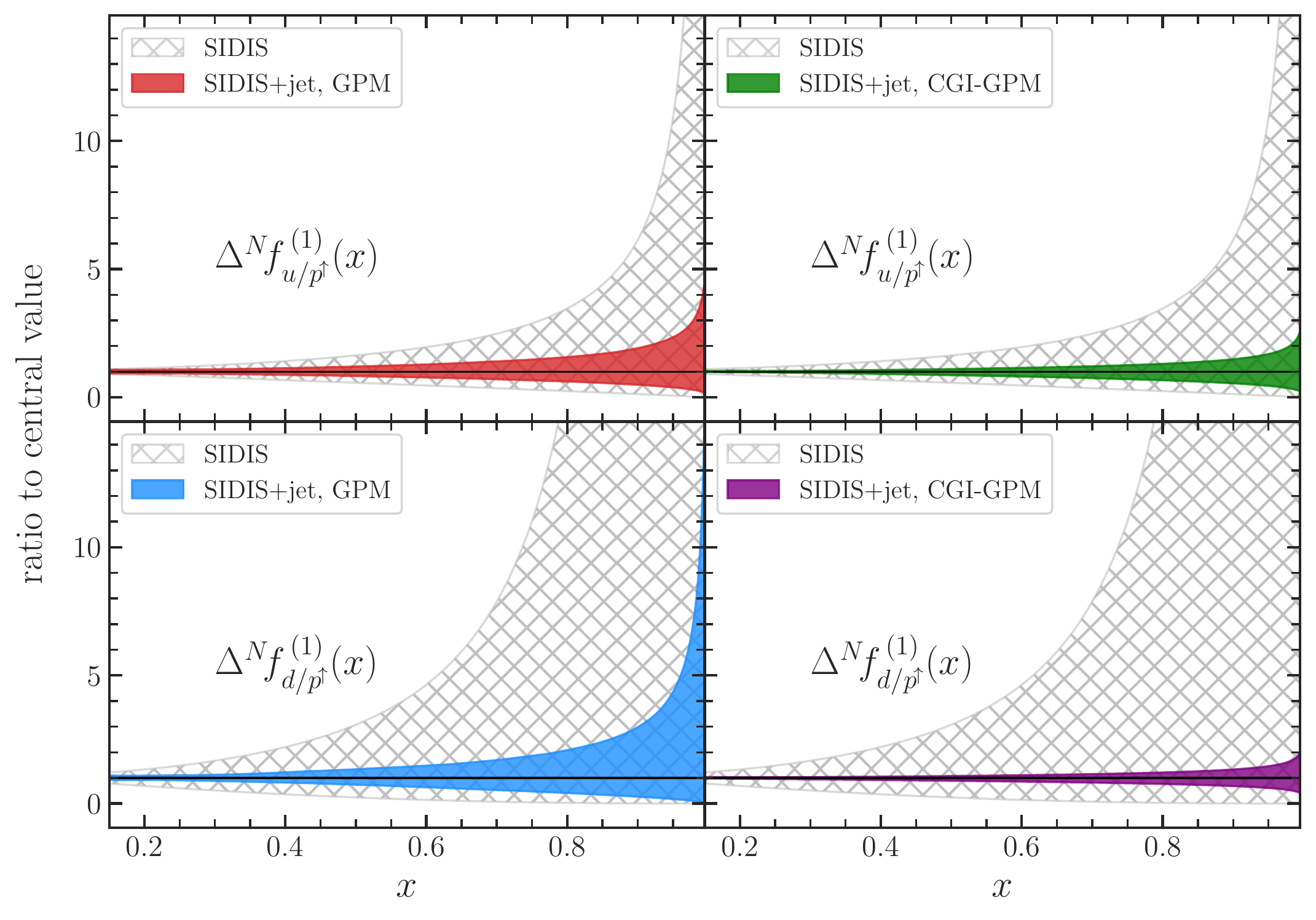}}\label{fig:siv-ratio}
\caption{Comparison between the Sivers first $k_\perp$-moments (a) and their values normalized to the corresponding central value (b) from SIDIS data and their reweighted SIDIS+jet counterparts in the GPM (left panels) and CGI-GPM (right panels) framework. In both plots, results for $u$- (upper panels) and $d$- (lower panels) quarks are shown. Bands correspond to a $2\sigma$ CL.}
\label{fig:Sivers}
\end{figure}

\begin{figure}[htp]
\centering
\includegraphics[width=\columnwidth]{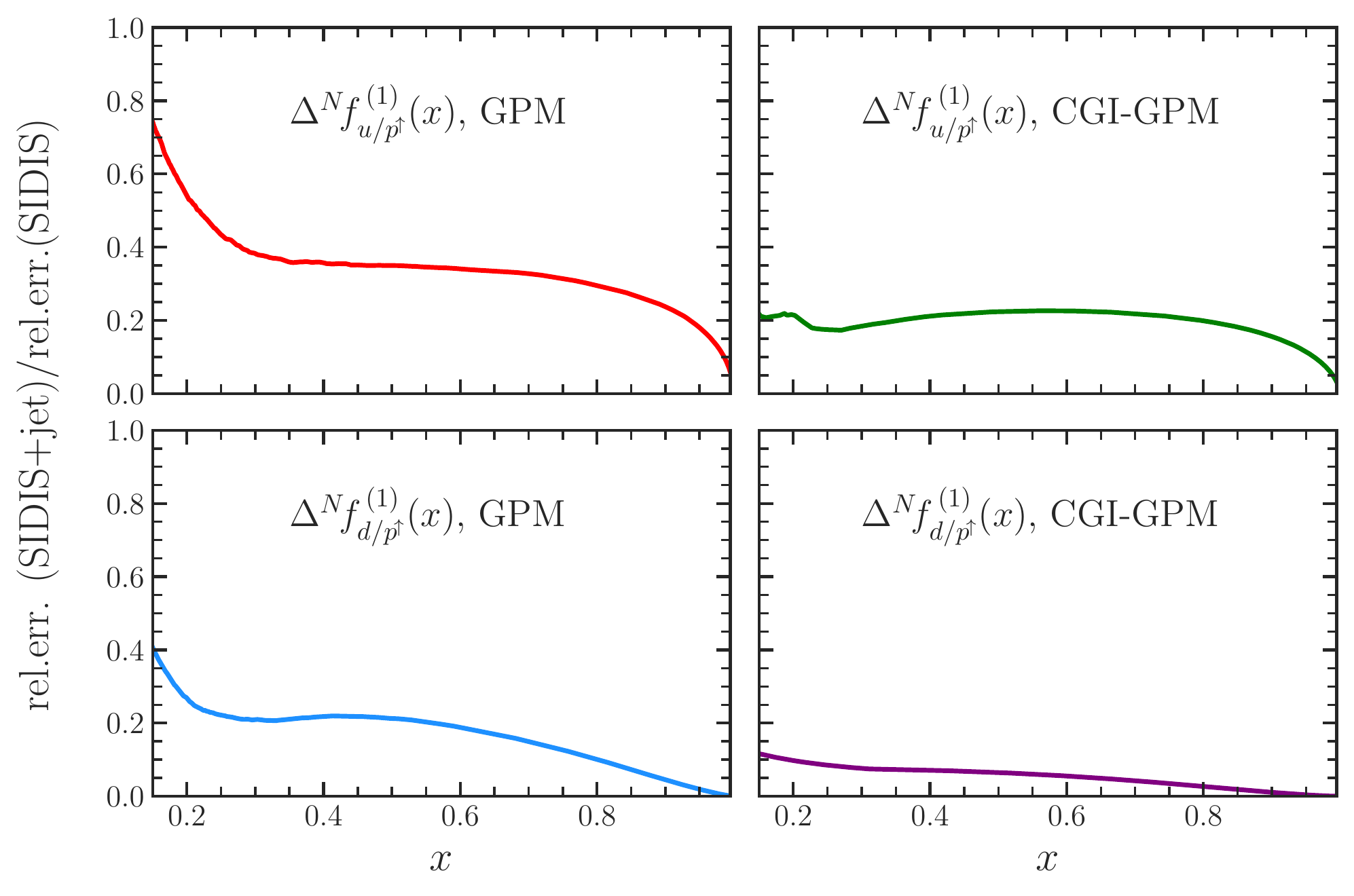}
\caption{Ratio of relative errors for the ``SIDIS+jet" and the ``SIDIS" cases in the GPM (left panels) and the CGI-GPM (right panels) approaches, for $u$-(upper panels) and $d$-(lower panels) quarks.}
\label{fig:Sivers-err-ratios}
\end{figure}

To better visualize the effect of the reweighting procedure on the Sivers function, in~\cref{fig:Sivers} we show the comparison between the prior Sivers first moments and the corresponding moments after reweighting in the GPM (left panels) and in the CGI-GPM (right panels) frameworks. More specifically, in Fig.~\ref{fig:Sivers}(a) we compare the first $k_\perp$-moments before and after the reweighting, while in Fig.~\ref{fig:Sivers}(b) we show the ratio of each Sivers first moment to its corresponding central value. As previously mentioned, the reweighting allows to constrain the Sivers function in the large-$x$ region, not covered by the current SIDIS data. No significant difference is observed in the low-$x$ region, as the model parametrization does not have any parameter controlling the low-$x$ behavior. The reduction of the uncertainty is dramatic, especially for the $d$-quark Sivers function in the CGI-GPM approach, that SIDIS leaves largely unconstrained. This is confirmed by the much narrower reweighted probability density for the $\beta_d$ parameter (see mid-right panel of~\cref{fig:pars-chi2ndof-distributions}).

In order to quantify the uncertainty reduction, in~\cref{fig:Sivers-err-ratios} we show the ratio between the relative errors on the Sivers function before and after the reweighting procedure, both for the GPM (left panels) and CGI-GPM (right panels) formalisms. In the GPM approach, we see an uncertainty reduction of about $60\%\,(80\%)$ for the reweighted $u$($d$)-quark Sivers function at $x > 0.2$, while in the CGI-GPM case, and in the same kinematical region, we have $\sim 80\%$ and $\sim 90\%$ for the $u$ and $d$ Sivers function, respectively. 

Before concluding this section, let us comment on the use of the jet data set, where no cut on the photon multiplicity is imposed. The corresponding results of the reweighting are indeed very similar to those already shown in all respects, apart from the fact that, in this case, the resulting $\chi^2_{\rm dof}$ slightly favors the CGI-GPM approach.

\section{\label{sec:conclusions}Conclusions and outlook}

In the present analysis we applied, for the first time, a reweighting procedure to a TMD parton density, the quark Sivers distribution function extracted from SIDIS data. By exploiting the recently published single spin asymmetry data for inclusive jet production in polarized $pp$ collisions from STAR~\cite{Adam:2020edg}, we showed the feasibility of such a procedure, which represents a valuable alternative to a full global fit.

This allowed us, for the first time, to combine  SIDIS azimuthal asymmetries data and the single spin asymmetries measured in $p^\uparrow p \to {\rm jet}\,X$ processes in a global analysis.
Moreover, by using two different approaches, the GPM and the CGI-GPM, we could also attempt to assess the degree of process dependence of the Sivers function, beside its sign change. Although the reweighted $\chi^2_{\rm dof}$ slightly favors the GPM approach, further investigations are needed to have a clear discrimination between the two formalisms.

By including the jet SSA data from STAR we were able  to significantly improve and extend our present knowledge of the quark Sivers function towards larger $x$ values, not probed in current SIDIS data. In particular, their high precision allows to remarkably reduce the uncertainties of the Sivers function  in such a kinematical region. We found that for the $u$-quark Sivers distribution, such reduction is about $60\%$ in the GPM and $80\%$ in the CGI-GPM frameworks, while for the $d$-quark case we observed a reduction of about $80\%$ and $90\%$ for GPM and CGI-GPM respectively.

This work has to be considered as an exploratory study to show, on one side, the potentiality of the reweighting procedure in the TMD sector and, on the other side, to refine the behavior of the Sivers function in a region not explored in SIDIS processes. 
The natural extension of this study will be a global analysis including also SSAs for inclusive pion production. This will allow us to simultaneously apply the reweighting procedure to the Collins fragmentation function, to transversity and to the Sivers function, as extracted by best-fitting the azimuthal asymmetries in SIDIS and $e^+e^-$ processes. 
We also expect that forthcoming SIDIS measurements from COMPASS~\cite{Bradamante:2018ick}, JLab~\cite{Dudek:2012vr} and the future Electron Ion Collider~\cite{Boer:2011fh,Accardi:2012qut} will play a crucial role in unravelling the nucleon structure in its full complexity. 

This rather ambitious program will provide important information on the impact of inclusive SSA data in the determination of these TMDs as well as a test of the compatibility of their extractions.

\section*{Acknowledgments}

We would like to thank the STAR Collaboration for providing us with the experimental data~\cite{Adam:2020edg}.
We are grateful to Mauro Anselmino for his involvement in the early stages of this work.
This project has received funding from the European Union's Horizon 2020 research and innovation programme under Grant No.~824093 (STRONG-2020) (M.B, U.D., C.F., J.O.G.H., F.M.),
by the National Science Foundation under the Contract  No.~PHY-2012002 (A.P.), and by the US Department of Energy under contract No.~DE-AC05-06OR23177 (A.P.) under which JSA, LLC operates Jefferson Lab, and within the framework of the TMD Topical Collaboration (A.P.).
C.F.~is thankful to the Physics Department of Cagliari University for the hospitality and support for his visit during which part of the project was done.

\section*{References}


\end{document}